# INVESTIGATING EFFORT PREDICTION OF SOFTWARE PROJECTS ON THE ISBSG DATASET

Sanaa Elyassami[1] and Ali Idri[1]

[1]Department of Software Engineering
ENSIAS, Mohammed Vth –Souissi University
BP. 713, Madinat Al Irfane, Rabat, Morocco

*ABSTRACT*

*Many cost estimation models have been proposed over the last three decades. In this study, we investigate fuzzy ID3 decision tree as a method for software effort estimation. Fuzzy ID software effort estimation model is designed by incorporating the principles of ID3 decision tree and the concepts of the fuzzy set-theoretic; permitting the model to handle uncertain and imprecise data when presenting the software projects.*

*MMRE (Mean Magnitude of Relative Error) and Pred(l) (Prediction at level l) are used, as measures of prediction accuracy, for this study. A series of experiments is reported using ISBSG software projects dataset. Fuzzy trees are grown using different fuzziness control thresholds.*
*Results showed that optimizing the fuzzy ID3 parameters can improve greatly the accuracy of the generated software cost estimate.*

## 1. INTRODUCTION

Estimation software project development effort remains a complicated problem, and one which continues to catch the attention of considerable research. Software projects expend 30%- 40% more effort than is estimated by the existing models [15].

Software development efforts estimation techniques may be grouped into two classes: parametric models [5], and non parametric models, which are based on a set of artificial intelligence methods such as case based reasoning [21], decision trees [22], artificial neural networks [4][11] and fuzzy logic [19][24].

There are two major types of software size metrics used to measure the effort that will be required to develop a project: Source Lines Of Code (SLOC) and Function Points (FP). The FP metric was initially developed as a substitute to SLOC to measure productivity in the later phases of software development. Though, the FP model demonstrates that it could also be a powerful to estimate software cost in the early phases of the software development lifecycle.

In this paper, we are concerned with a non-parametric model witch is based on fuzzy decision tree and which the effort is measured by using the FP software size metric rather than SLOC metric.

  

The decision tree technique is usually used for inductive learning and has been signifying its superiority in terms of predictive accuracy in several fields [12] [25]. The most widely used algorithms for building a decision tree are ID3 [13], C4.5 [14] and CART [16].

The most important benefits when using estimation by decision trees (DT) are: First, decision trees approach may be considered as "white boxes", it is simple to understand and easy to explain its process to the users, contrary to other learning methods. Second, DT may be used to feature subset selection to exceed the problem of cost driver selection in software cost estimation model. Thirdly, Decision tree permits the learning from preceding situations and outcomes.
On the other hand, we have used fuzzy logic in software cost estimation. The fuzzy set theory was introduced by Zadeh in 1965 [17].

Several attempts have been made to revitalize some of the existing models in order to handle imprecision problems and uncertainties. Idri et al. [3] studied the application of fuzzy logic to the cost drivers of intermediate COCOMO model, while Pedrycz et al. [26] investigate a fuzzy set approach to estimate software project efforts.

In earlier works [1] [2] we have empirically evaluated the use of the crisp decision tree and the fuzzy decision tree techniques for software cost estimation. The used decision tree algorithms are the ID3, the C4.5 and the fuzzy ID3. The previous studies are based on the COCOMO' 81 and on the web hypermedia Tukutuku dataset. From the obtained results we have found that fuzzy decision trees permit to exploit advantages of fuzzy logic theory which is the ability to deal with inexact and uncertain information when describing the software projects.

In this work, we are concerned with cost estimation models based on Fuzzy Interactive Dichotomizer 3 decision tree and applied on the ISBSG dataset which the effort is measured by using the FP software size metric.

The remainder of this paper is organized as follows: In section II, we present the fuzzy ID3 decision tree for software cost estimation. The description of dataset used to perform the empirical studies and the data preprocessing are given in section III. Section IV focuses on the experimental design and the evaluation criteria adopted to measure the predictive accuracy of the model. In Section V, we present and discuss the obtained results when the fuzzy ID3 is used to estimate the software effort.

## 2. FUZZY ID3 FOR SOFTWARE COST ESTIMATION

The fuzzy ID3 is founded on a fuzzy implementation of the Interactive Dichotomizer 3 (ID3) algorithm [18] [23]. The major characteristic of fuzzy ID3 is that an example belongs to a node to a certain degree. The proportion of $p_k^n$ of examples with $k$ classification at node $n$ is calculated using the membership degrees as follows:

$$p_k^n = \frac{\sum_{i=1}^{N} u_k(y_i) \wedge u_n(x_i)}{\sum_{c=1}^{K} \sum_{i=1}^{N} u_c(y_i) \wedge u_n(x_i)} \qquad (1)$$

International Journal of Artificial Intelligence & Applications (IJAIA), Vol.3, No.2, March 2012

Where K represents the classes and N is the number of examples in the subset. $u_k(y_i)$ is the membership degree on the web project $i$ that belongs to the class $k$ and $u_n(x_i)$ is the membership degree of the web project $i$ at node $n$.

$\wedge$ represents the conjunction operator. T-norm, which generalizes intersection in the domain of fuzzy sets, is usually used for fuzzy conjunction. The most popular T-norms are minimum and product.

The fuzzy entropy uses the membership degree of examples at a particular node and contributes to enhance the discriminative power of an attribute, is computed as:

$$H_n = -\sum_{k} p_k^n * \log(p_k^n) \qquad (2)$$

The expansion of the fuzzy ID3 is realized by growing a node of tree categorized by the highest information gain. The information gain is calculated as:

$$G_n^j = H_n - \sum_{l=1}^{M} w_l H_l \qquad (3)$$

Where $H_n$ is the entropy in the node $n$. $H_l$ is the entropy of the node that belongs to the fuzzy set $L$ of the $j$ variable. $w_l$ is the fuzzy set relative weight.

The algorithm terminated as follow:

- If the classification truth degree of a branch with respect to one class exceeds a specified threshold, the branch is finished as a leaf
- At a branch, all attributes are used for splits, the branch is terminated as a leaf or as null
- At a node, all examples have the same classification
- If one of the conditions mentioned above is satisfied, the expanding of the branch in a fuzzy decision tree will be terminated.

## 3. ISBSG REPOSITORY

The International Software Benchmarking Standards Group (ISBSG) data repository was used in this study. The ISBSG repository (release 8) consists of 2027 projects collected from twenty countries around the world. Major contributors are Australia (21%), Japan (20%) and the United States (18%).

The reduction of dimensionality in ISBSG dataset is primordial to operate faster and improve classification accuracy.

This section describes the preprocessing carried out on the ISBSG dataset. To do so, we performed data transformation and data selection.

### 3.1. Data Transformation

Data transformation enables algorithms to be applied without difficulty and improves their performance and their effectiveness. Data transformation operations contribute to get the required information from incomplete, noisy and incoherent set of data. We have chosen in this study to applied two data transformation operations:

- Handle missing values
- Data normalization

### 3.1.1. Missing Data

Missing Data is the common problem that comes up through the data preparation stage. There are many approaches to deal with missing values, for instance: Avoid Missing data and Data Imputation.

The first approach generally lost too much useful information. In this study, the second method is chosen to handle missing values.

Imputation methods use information available in the dataset to predict missing values. These methods may be grouped into two categories: single and multiple imputation methods [6]. In single imputation methods, the missing value is replaced with one imputed value, and in multiple imputation methods, several values are used.

To handle missing data in ISBSG dataset, the Mean or Mode Single Imputation (MMSI) method is used. For continuous attributes, MMSI method substitutes the missing value with the mean. For nominal attributes, MMSI method substitutes the missing value with the mode (the value that is repeated more often than any other).

For example, for the language type attribute, the dominant value in the dataset is 3GL (67%). So, in the case of language type attribute, the missing values are replaced by 3GL.

### 3.1.3. Normalizations

Due to the nature of software attributes, some of continuous features show a larger range of values than others which may make the effect of this feature too important or easily neglected. The solution is to scale continuous features into the same range. To achieve this, all continuous features are normalized applying the min max normalization formula as show in eq. (4) such that all numeric variables are scaled in the range [0, 1].

$$x_i(k) = \frac{x_i(k) - \min(x(k))}{\max(x(k)) - \min(x(k))} \quad (4)$$

## 3.2. Data Selection

To increase the efficiency of prediction accuracy by FDT we have to apply some data selection techniques to obtain a reduced representation of the data set that is much smaller in volume, yet closely maintains the integrity of the original data.
In this study, two operations in data reduction are used:

- Feature Selection
- Case Selection

### 3.2.1. Feature Selection

The aim is to identify the features that have the highest potential to provide good effort estimates. A function point is used as a measure of software size and was chosen in this study, firstly. The second variable is team size that is a potential cost factor. Thirdly, development platform was chosen, where each project is classified as either, a PC, Mid Range or Main Frame. Other important criterion for selecting projects is user base. User base enclose 3 criteria: the first one is the number of business units that the system services, the second one is the number of physical locations being serviced by the system and the last one is the number of users using the system concurrently.

Finally, the project effort is used as the dependent feature that provides the total effort for all phases of the project development life cycle.

Seven criteria are chosen in estimation data set, these criteria are also suggested by ISBSG. Table I includes the project metrics that have been used in this study.

Table 1. Project metrics used

| Metric | Definition |
| --- | --- |
| Function Points | adjusted function point count number |
| Max Team Size | maximum number of people on the project |
| User Base - Business Units | number of business units that the system services |
| User Base - Locations | number of physical locations being serviced by the system |
| User Base - Concurrent Users | number of users using the system concurrently |
| Development Platform | primary platform (PC, Mid Range or Mainframe) |
| Normalized Work Effort | total effort in hours recorded against the project for all phases of the development life cycle |

### 3.2.2. Case Selection

In order to obtain a reduced data set, some projects had to be excluded. The raw data was filtered by several criteria. Four filtering criteria are used in this study:

- Data Quality Rating
- Resource Levels
- Rating for Unadjusted Function Points
- Development Type

#### 3.2.2.1. Data Quality Rating

We have to ensure that the model is built on the basis of a reliable data set. According to ISBSG, only projects with data quality ratings A or B were included. Projects rated C and D offer valuable data, but uncertainty about some of their size or effort values. Hence, projects whose quality ratings are C and D were not included in the estimation data set.

#### 3.2.2.2. Resource Levels

Four resource levels are identified in the ISBSG data collection instrument:

1 = development team effort

2 = development team support
3 = computer operations involvement
4 = end users or clients

Generally, the cost estimation models take into account only development team effort and support (resource level 1 and resource level 2) rather than considering the other costs like computer operations involvement and effort expended by end user (resource level 3 and resource level 4). Therefore, only projects recorded at the first resource level or at the second resource level were considered. So, the work effort for the development team and support is included in the work effort number. Projects with level 3 or level 4 were discarded. The goal is to make the results as generalizable as possible.

### 3.2.2.3. Rating for Unadjusted Function Points

Rating for Unadjusted Function Points (UFP) is an ISBSG rating code applied to the unadjusted FP data by the ISBSG quality reviewer to measure the UFP integrity. Projects of UFP A or B are included in the subset. Projects whose UFP ratings are C and D were left.

### 3.2.2.4. Development Type

Development type is the final criterion on which we based our investigation. This measure describes whether the development is a new development, enhancement or redevelopment. Only the new development projects are considered in our study.

Table 2. The criteria used to reduce the dataset

| Criteria | Selected Data | Discarded Data |
|---|---|---|
| Development Type | New Development | Enhancement and Redevelopment |
| UFP | A or B | C and D |
| Resource Levels | 1 or 2 | 3 and 4 |
| Data Quality Rating | A or B | C and D |

Table 2 summarize, the original data set was reduced as follows:

- Remove projects if they were not assigned a high data quality rating (A or B) by ISBSG.
- Remove projects with resource levels different from 1 or 2 (development team effort and development team support only).
- Remove projects if their rating for unadjusted function points different from A or B.
- Remove projects with development type enhancement or redevelopment and keep only new developments projects.

The application of all of these criteria results in an important decrease in the number of projects. Only 151 projects left after applying the filtering criteria.

## 4. EXPERIMENT DESIGN

For each experiment, the original data set (151 projects) was randomly separated into 74 training data projects and 77 test data projects.

We applied the FID algorithm for the induction of the decision trees. We have used the latest release of FID3.4 [7]. The FID 3.4 is available at http://www.cs.umsl.edu/~janikow/fid/.
The use of fuzzy ID3 algorithm to estimate software development effort requires the determination of the following parameters:

- T-norms
- Inference method
- Fuzzy discretization
- Stop criteria

All these parameters need to be optimized. Several values were tested and the optimal ones were used.

In the present paper we are interested in studying the impact of thresholds controlling the growth of the tree on the accuracy of fuzzy ID3 and especially the fuzziness control threshold.

### 4.1. T-norms

Referring to equation 1, we have to specify the t-norm operator that will be used to calculate the fuzzy entropy during tree building. Noting that minimum T-norm and product T-norm are the two commonly used fuzzy conjunction operators because of their well behaviour and their computational simplicity [10].

We have conducted two preliminary experiments to decide witch operator we have to choose.
The first fuzzy ID3 effort estimation model uses the product conjunction operator and the second fuzzy ID3 effort estimation model uses the minimum conjunction operator.

The results show that product t-norm perform much better in terms of predictive accuracy that the minimum t-norm. So, the product T-norm was used for all the experiments.

### 4.2. Inference method

Once a fuzzy decision tree is generated for the training data, the tree can be used to classify new projects. To determine the class of new projects, FID3.4 performs two different methods of inference: set-based and exemplar-based [8]. Each has several manners of resolving internal conflicts (Leaf containing training data from multiple classes) and external conflicts (Multiple leaf activations with different degree of match).

Set-based inferences treat leaves as fuzzy sets, while exemplar-based inferences treat leaves as super-training exemplars. For all the experiments, the exemplar-based inference was used.

### 4.3. Fuzzy Discretisation

Each ISBSG software project is described by a set of selected attributes which can be measured by numerical or categorical values. Building a fuzzy ID3 decision tree require a fuzzy discretization of numeric attributes. So, these values will be represented by fuzzy sets.

The cost drivers and work effort are discretized into trapezoidal fuzzy sets. Fuzzy sets are automatically generated by using the local discretization [9] implemented in FID 3.4.

For example, in the case of the MTS (Max Team Size) cost driver, a fuzzy set for each linguistic value was defined with a trapezoidal membership function. For the other cost drivers, we proceed in the same way as for MTS. thus; the fuzzy sets corresponding to the various associated linguistic values were defined for each cost driver. The number of fuzzy partitions created for each cost driver is presented in table 3.

In the case of the work effort, it was represented with both 11 fuzzy sets and with 16 fuzzy sets. The aim is to test which configuration can improve results. Details are given in the next section.

Table 3. The number of fuzzy sets used per attribute

| Input Variable | Attribute type | Number of Fuzzy sets |
|---|---|---|
| Function Points | Linear | 7 |
| Max Team Size | Linear | 11 |
| User Base - Business Units | Linear | 9 |
| User Base - Locations | Linear | 9 |
| User Base - Concurrent Users | Linear | 9 |
| Development Platform | Nominal | - |
| Work Effort | Linear | 11/16 |

### 4.4. Stop Criteria

In building a fuzzy decision tree, without threshold controlling the growth of the tree, the fuzzy decision tree generated will be very large and complicated. The fuzzy decision tree cannot be overly large because of the limited computing capacity.

Different stop criteria are provided in FID:

- The fuzziness control threshold verifies if the ratio of membership of a class at tree node is higher than a given threshold
- Information content at which expansion should be stopped
- The leaf decision threshold verifies if the number of objects at tree node is less than a given threshold

The stop criterion used is the fuzziness control threshold that takes on continuous values between 0 and 1. To optimize this parameter for the fuzzy ID3 model, the value for the fuzziness control threshold was varied between 0.1 and 0.9. The results were given in table IV.

The fuzziness control threshold Th controls the growth of the fuzzy tree. Different Th will generate different fuzzy ID3 decision tree with different classification accuracy, lower Th may lead to a larger tree, but the classification accuracy will be higher. High Th may lead to a small tree while the classification accuracy may be lower.

### 4.5. Evaluation Criteria

For the purpose of evaluation and validation, it is necessary to measure how accurate the software estimates are. We employ the subsequent criteria to compute the accuracy of the estimates

128

produced by the fuzzy ID3. A common criterion for the evaluation of effort estimation models is the magnitude of relative error (MRE), witch is defined as:

$$MRE = \left| \frac{Effort_{actual} - Effort_{estimated}}{Effort_{actual}} \right| \qquad (5)$$

where $Effort_{actual}$ is the actual effort of a project in the dataset, and $Effort_{estimated}$ is the estimated effort that was obtained using a model or a technique.

The MRE values are measured for each project in the datasets, while mean magnitude of relative error (MMRE) computes the average over N projects.

$$MMRE = \frac{1}{N} \sum_{i=1}^{N} \left| \frac{Effort_{actual,i} - Effort_{estimated,i}}{Effort_{actual,i}} \right| \times 100 \qquad (6)$$

The acceptable target values for MMRE are $MMRE \leq 25$.

The prediction Pred("p") was also used. It's the percentage of MRE that is less than or equal to the value "p" among all projects [20] the used value for p is 25%.

The definition of Pred(p) is given as follows:

$$Pred(p) = \frac{k}{N} \qquad (7)$$

Where k is the number of observations whose MRE is less or equal to p and N is the number of the total observations. The acceptable values for Pred(25) are $Pred(25) \geq 75$.

## 5. OVERVIEW OF THE EXPERIMENTAL RESULTS

We conducted several experiments for each combination of the parameters (the number of work effort fuzzy sets and the fuzziness control threshold value). The aim is to determine which configuration improves the estimates.

Two models were generated. The first model was generated using 11 fuzzy sets to represent work effort (11 FS model) and the second one was generated using 16 fuzzy sets to represent work effort (16 FS model).

For each model, the fuzziness control threshold was varied between 0.1 and 0.9. The verification results for the two models are shown in Table IV.

Table 4. The verification results of 11-class and 16-class models

| Threshold value | 11 work effort fuzzy sets | | 16 work effort fuzzy sets | |
|---|---|---|---|---|
| | MMRE | Pred(25) | MMRE | Pred(25) |
| 0.1 | 15,41 | 84,15 | 13,49 | 92,2 |
| 0.2 | 21,31 | 83,16 | 14,02 | 89,61 |
| 0.3 | 23,21 | 77,92 | 15,58 | 85,71 |
| 0.4 | 28,57 | 68,83 | 16,25 | 83,16 |
| 0.5 | 39,47 | 53,24 | 23,98 | 75,32 |
| 0.6 | 48,37 | 45,45 | 33,73 | 68,83 |
| 0.7 | 64,8 | 23,37 | 38,94 | 53,24 |
| 0.8 | 74,15 | 18,18 | 49,32 | 45,45 |
| 0.9 | 64,82 | 15,58 | 52,68 | 23,37 |

The results of MMRE of the software cost estimation model using the FID fuzzy decision tree are shown in Fig. 1. In general, the 16 FS model has lower MMRE than the 11 FS model. The 16 FS model has an acceptable MMRE for the fuzziness control threshold values 0.1, 0.2, 0.3, 0.4 and 0.5.

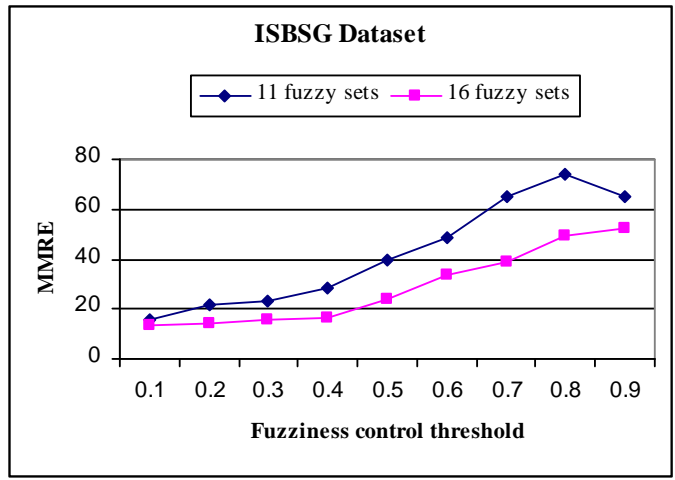

Figure 1. The MMRE in 11 FS and 16 FS models

Fig. 2 presents the results of Pred(25) when using the FID fuzzy decision tree.

The 16 FS model has higher Pred(25) than the 11 FS model. The 16 FS model has an acceptable Pred(25) for the fuzziness control threshold values 0.1, 0.2, 0.3, 0.4 and 0.5.

For 16 FS model with fuzziness control threshold value equals to 0.4, more than 83.16% predicted values of project effort falling within 25% of their actual values.

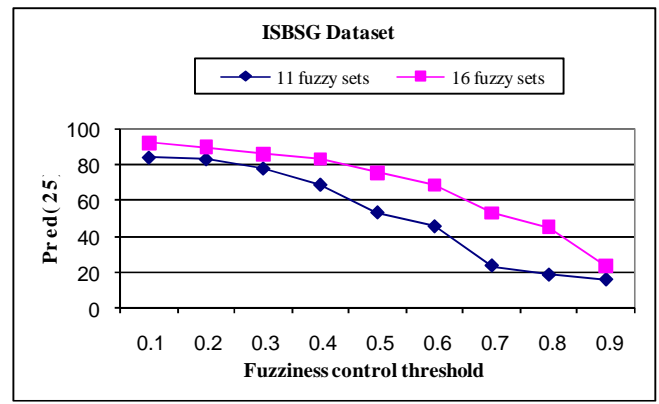

Figure 2. The Pred(25) in 11 FS and 16 FS models

Different threshold generate different fuzzy ID3 decision tree with different classification accuracy. Table 4 shows clearly that the accuracy of the estimates generated by the fuzzy ID3 decreases with the growth of the fuzziness control threshold value. The fuzziness control threshold growth may lead to a small tree that explains the decrease of the classification accuracy. Thus demonstrating that optimizing the fuzzy ID3 parameters can improve the accuracy of the generated software cost estimates.

## 6. CONCLUSION

Fuzzy ID3-based model is trained and tested using the ISBSG software projects dataset in this paper.

Data preprocessing operations have been performed on the ISBSG reducing the dataset dimensionality and enabling fuzzy ID3 decision tree algorithm to operate faster and without difficulty.

Experiments were conducted with different thresholds to control the growth of trees. Further, the effect of changes in the number of fuzzy sets in the model is investigated. The results show that preprocessing of input data improves the performance of the applied algorithm. Furthermore, optimizing fuzzy ID3 parameters improves greatly the generated software effort estimates.

## Authors

**S. Elyassami** received her engineering degree in Computer Science from the UTBM, Belfort-Montbeliard, France, in 2006. Currently, she is preparing her Ph.D. in computer science in ENSIAS. Her research interests include software cost estimation, software metrics, fuzzy logic and decision trees.

**A. Idri** is a Professor at Computer Science and Systems Analysis School (ENSIAS, Rabat, Morocco). He received DEA (Master) (1994) and Doctorate of 3rd Cycle (1997) degrees in Computer Science, both from the University Mohamed V of Rabat. He has received his Ph.D. (2003) in Cognitive Computer Sciences from ETS, University of Quebec at Montreal. His research interests include software cost estimation, software metrics, fuzzy logic, neural networks, genetic algorithms and information sciences.